\documentclass[showpacs,aps,prd,longbibliography,nofootinbib,showkeys,superscriptaddress,twocolumn,longbibliography]{revtex4-2}
\usepackage{graphicx}
\usepackage{bm}
\usepackage{multirow}
\usepackage{natbib}
\usepackage{array}
\usepackage{amsmath}
\usepackage{mathrsfs}
\usepackage{amssymb}
\usepackage{float}
\usepackage{xcolor}
\usepackage{enumerate}
\usepackage{color}
\usepackage{slashed}
\usepackage{hyperref}
\usepackage{url}
\usepackage{txfonts}
\usepackage{mathtools}
\usepackage{booktabs}
\usepackage{BOONDOX-cal}
\usepackage{orcidlink}
\usepackage{longtable}
\usepackage{makecell}
\usepackage{cancel}

\allowdisplaybreaks[4]
\hypersetup{hypertex=true,colorlinks=true,linkcolor=blue,anchorcolor=red,citecolor=blue}

\begin{document}
	
	\title{Spectra and strong decays of open charm-bottom $c\bar{b}q\bar{q}$ tetraquarks in an unequal-mass Born-Oppenheimer framework}
	
	\author{Bowen Kang}
	\affiliation{College of Physics, Chengdu University of Technology, Chengdu 610059, China}
	
	\author{Xi Xia}
	\affiliation{College of Physics, Chengdu University of Technology, Chengdu 610059, China}
	
	\author{Tao Guo}
	\email[Corresponding author: ]{guot17@tsinghua.org.cn}
	\affiliation{College of Physics, Chengdu University of Technology, Chengdu 610059, China}
	
	\date{\today}
	
\begin{abstract}

We investigate the mass spectra and strong decay channels of $S$-wave open charm-bottom $c\bar{b}q\bar{q}$\ ($q = u,d,s$) tetraquarks within the Born-Oppenheimer approximation. Extending our previous study of hidden heavy flavor tetraquarks, we account for the unequal-mass heavy quark sources and construct the effective potential based on independently optimized heavy-light orbitals. This makes the computational method more generally applicable. For the quantum numbers $J^P=0^+$, $1^+$, and $2^+$, we have obtained the spectra of the open charm-bottom $c\bar{b}q\bar{q}$ tetraquarks corresponding to color-singlet states constructed by coupling two color octets. The predicted masses exhibit systematic flavor-dependent shifts and characteristic hyperfine splittings. We also estimate partial widths for selected two-body open flavor decay channels using color-spin recoupling and a phenomenological decay prescription. The resulting channel-dependent decay modes, together with comparisons to other theoretical predictions, provide useful benchmarks for future theoretical investigations and experimental searches for open charm-bottom tetraquarks.

\end{abstract}

\maketitle

\section{Introduction}

Quantum chromodynamics (QCD) is the established theory of the strong interaction and has successfully described the spectra of conventional hadrons within the constituent quark model~\cite{Godfrey:1985xj,Capstick:1986ter}, in which mesons and baryons are identified as quark-antiquark and three-quark bound states, respectively~\cite{Gell-Mann:1964ewy}. 
Historically, this model provided an economical and reliable framework for classifying hadrons, until the discovery of the narrow charmonium-like state $X(3872)$ by the Belle collaboration in 2003~\cite{Belle:2003nnu}.
Since then, a large number of unconventional hadrons (i.e., exotic hadrons), collectively referred to as the $XYZ$ states, have been observed by various experimental collaborations~\cite{Brambilla:2019esw,Chen:2022asf,ParticleDataGroup:2024cfk,Liu:2024uxn}. 
Many of these exotic hadrons exhibit properties that are difficult to reconcile with conventional quark-antiquark or three-quark configurations, indicating the presence of novel QCD dynamics beyond the traditional quark model.
At the same time, the emergence of these hadrons has substantially enriched the landscape of the hadronic system and  provides an important testing ground for our understanding of nonperturbative QCD.

A variety of theoretical models have been proposed to explain the nature of these exotic hadrons.
Among the most extensively studied scenarios are hadronic molecules~\cite{Janssen:1994wn,Shmatikov:1995aj,Guo:2017jvc}, compact multiquarks~\cite{Jaffe:1976ig,Esposito:2016noz,Ali:2017jda,Guo:2021yws,Xia:2025mgk}, glueballs~\cite{Close:1998mj,Brunner:2015yha,Fariborz:2025ybo}, and hybrids~\cite{Chanowitz:1982qj,Klempt:2007cp,Meyer:2015eta}.
Based on our previous research on hidden heavy flavor $Q\bar{Q}q\bar{q}$ ($Q = c,b$) tetraquark states~\cite{Kang:2025xqm}, we focus our attention on asymmetric open heavy flavor $c\bar{b}q\bar{q}$ tetraquark system.
This $c\bar{b}q\bar{q}$ tetraquark system contains both heavy and light quark components, and the large mass hierarchy difference between the heavy and light quarks naturally leads to a separation of dynamical scales.
In close analogy with the hydrogen molecule in molecular physics, the slow motion of heavy quarks can be treated independently from the fast dynamics of the light quarks. 
This idea motivates the application of the Born-Oppenheimer (BO) approximation, which has proven highly successful in quantum electrodynamics, especially in the study of molecular binding and dissociation phenomena~\cite{Piszczatowski2009,Puchalski:2018pdf,Silkowski:2023mjy}. 
Currently, significant progress has been made in understanding the structure and dynamics of hadrons using the BO approximation~\cite{Fleck:1989mb,Braaten:2014qka,Brambilla:2017uyf,Bicudo:2017szl,Maiani:2019lpu,Germani:2020sgm,Berwein:2024ztx,Liu:2025jyn}. 
Given the analogous hierarchy in hadron systems composed of heavy and light quarks, extending this approach to QCD is obviously both natural and theoretically well justified.

Experimentally, the major heavy flavor tetraquark candidates observed so far all contain hidden-charm $c\bar{c}$ or hidden-bottom $b\bar{b}$ components~\cite{Chen:2016qju,Liu:2019zoy,CMS:2025fpt}.
Nevertheless, the mystery of open heavy flavor tetraquark states has sparked the curiosity of physicists.
In 2016, the D0 collaboration reported an open-bottom exotic hadronic state $X(5568)$, whose components were considered to be tetraquarks $b\bar{s}u\bar{d}$, $b\bar{s}d\bar{u}$, $s\bar{b}u\bar{d}$, or $s\bar{b}d\bar{u}$~\cite{D0:2016mwd}.
Unfortunately, subsequent experiments did not find strong evidence for the existence of this state, such as the ATLAS~\cite{ATLAS:2018udc}, LHCb~\cite{LHCb:2016dxl}, CDF~\cite{CDF:2017dwr}, and CMS~\cite{CMS:2017hfy} collaborations.
In 2020, the LHCb collaboration reported two open-charm exotic hadronic states $X_{0,1}(2900)$ in the $B^+\rightarrow D^+D^-K^+$ decay, and their valence quark contents were thought to be the $u\bar{c}d\bar{s}$ or $c\bar{d}u\bar{s}$ tetraquarks~\cite{LHCb:2020bls,LHCb:2020pxc}.
This discovery has once again inspired experimental research on open heavy exotic hadronic states.
Recently, the first observations of an open-charm doubly charged tetraquark state $X_{T_{c\bar{s}0}}(2900)^{++}$ and its neutral partner $X_{T_{c\bar{s}0}}(2900)^0$ were achieved by the LHCb collaboration in the invariant mass spectra of $D_s^+\pi^+$ and $D_s^+\pi^-$~\cite{LHCb:2022sfr,LHCb:2022lzp}, respectively.
Clearly, the discovery of more possible open heavy flavor tetraquark states is ongoing.

For the lighter $c\bar{s}q\bar{q}$ tetraquark system, theoretical calculations agree well with experimental observations~\cite{Liu:2020nil,Zhang:2020oze,Wang:2020prk,Guo:2021mja,Duan:2023qsg,Wang:2023hpp}. 
For the heavier $c\bar{b}q\bar{q}$ tetraquark system, reliable experimental observations are still lacking, but this has nevertheless attracted considerable theoretical attention.
Their mass spectra and decay modes have been explored using constituent quark models~\cite{Ebert:2007rn,Ortega:2020uvc}, chromomagnetic interaction model~\cite{Wu:2018xdi}, Regge trajectory analyses~\cite{Chen:2025fyh}, QCD sum rules~\cite{Wang:2019xzt,Agaev:2016dsg}, and chiral effective field theory~\cite{Liu:2025fhl}. 
In this work, we further investigate the asymmetric open charm-bottom $c\bar{b}q\bar{q}$ tetraquark system.
Within the diquark-antidiquark picture, we employ the BO approximation to construct effective potentials incorporating both color-Coulomb and color-spin hyperfine interactions.
By solving the Schr{\"o}dinger equation associated with light quarks, we determine the BO effective potential, and subsequently solve the resulting two-body Schr{\"o}dinger equation for heavy quarks with unequal masses.
Furthermore, we calculated the spectra of the tetraquark $c\bar{b}q\bar{q}$ system and analyzed their dominant two-body decay modes.

This paper is organized as follows. 
In Sec. \ref{S2}, we first give the Hamiltonian of the $c\bar{b}q\bar{q}$ tetraquark system.
Then we solve the BO effective potential of the system and construct the corresponding color-spin-flavor basis wavefunctions.
In addition, a phenomenological formula for calculating partial decay widths is presented.
In Sec. \ref{S3}, the method for extracting model parameters is described, and the required model parameters are listed.
We then primarily discuss and analyze our theoretical findings.
Finally, in Sec. \ref{S4}, we summarize our work and reiterate some important findings. 
Nature units $c = \hbar = 1$ are used throughout.

\section{Theoretical approach \label{S2}}

In our previous study, we applied the BO approximation to hidden-heavy tetraquark systems with equal heavy-flavor constituents. 
Here we generalize that construction to the flavor-asymmetric open charm-bottom $c\bar b q\bar q$ tetraquark system. 
In particular, The unequal charm- and bottom-quark masses require a separate treatment of the two heavy sources and the use of the heavy-pair reduced mass in the adiabatic Schr\"odinger equation. 
Throughout this work, the states we investigate are limited to $S$-wave tetraquarks. 
Consequently, their positive parity arises from the intrinsic parities of the two quarks and two antiquarks.
Since $c\bar b q\bar q$ is not an eigenstate of charge conjugation, these states are labeled solely by the quantum numbers $J^P$.

At the level of the mass spectrum of tetraquark, we express the effective Hamiltonian as~\cite{Kang:2025xqm}
\begin{equation}
	H_{\rm eff}=\sum_{i=1}^{4}m_i+H_{\rm BO}+H_{ss},
	\label{eq1}
\end{equation}
where $m_i$ is the mass of the constituent quark, $H_{\rm BO}$ describes the spin-independent adiabatic dynamics, and
\begin{equation}
	H_{ss}=\sum_{i<j}2\kappa_{ij}\,\bm s_i\cdot\bm s_j
	\label{eq2}
\end{equation}
is the effective color-spin hyperfine interaction. 
The coefficients $\kappa_{ij}$ absorb the short-distance spatial overlap, the inverse constituent-mass dependence, the running strong coupling constant, and the color matrix element of the corresponding quark pair. 
Their numerical determination is discussed in Sec.~\ref{S3}.

\subsection{Born-Oppenheimer reduction for unequal heavy masses}

\begin{figure}[htbp]
	\centering
	\includegraphics[width=0.42\columnwidth,keepaspectratio]{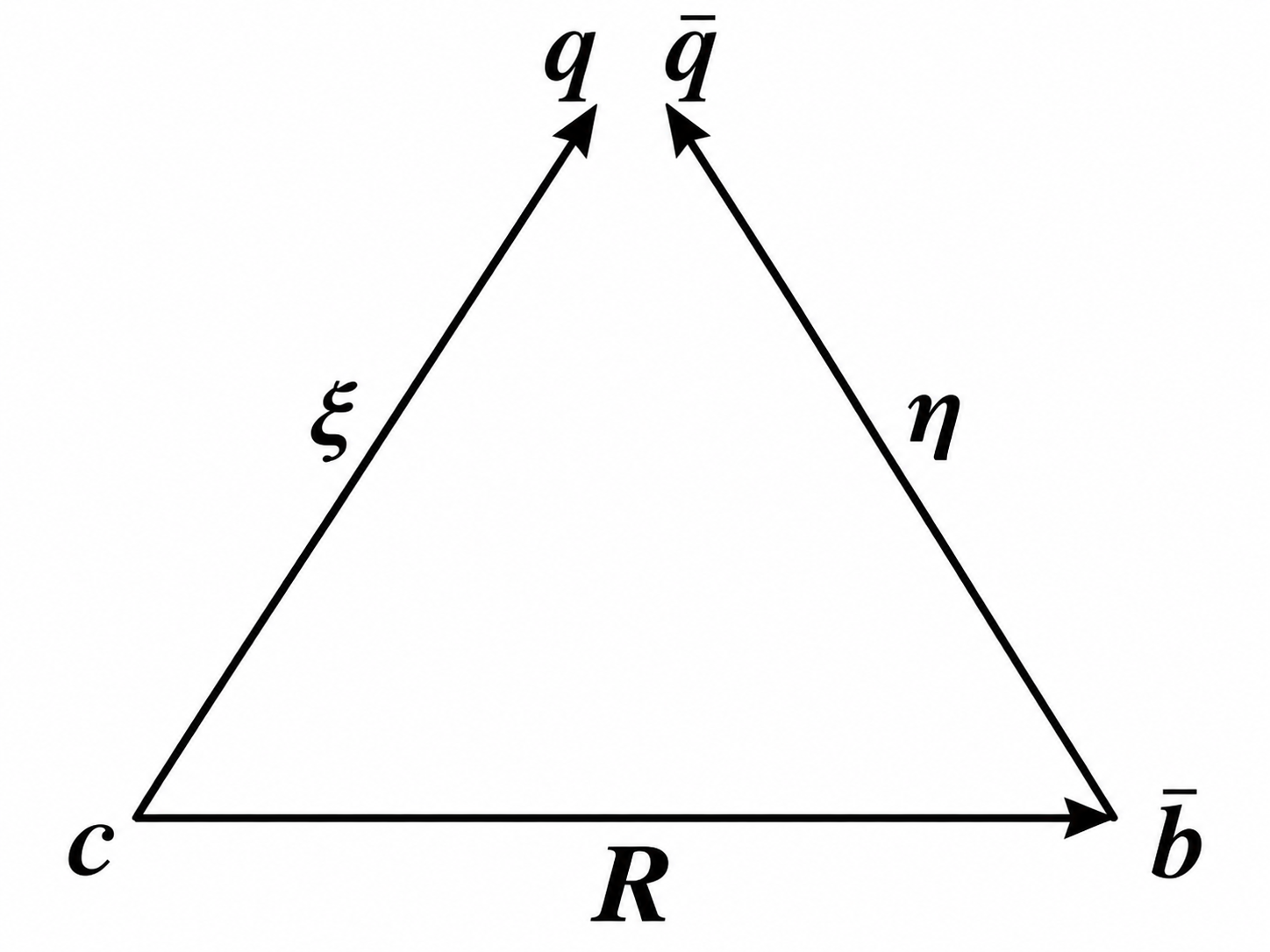}
	\caption{Coordinates used in the BO construction. The heavy sources $c$ and $\bar b$ are separated by $\bm R=\bm x_B-\bm x_A$, while $\bm\xi=\bm x_1-\bm x_A$ and $\bm\eta=\bm x_2-\bm x_B$ locate the light quark $q$ and antiquark $\bar q$ relative to the corresponding heavy source.}
	\label{fig:vector_diagram}
\end{figure}

We denote the positions of the charm and anti-bottom quarks by $\bm x_A$ and $\bm x_B$, respectively, and the positions of the light quark and antiquark by $\bm x_1$ and $\bm x_2$. 
The heavy-pair separation reads
\begin{equation}
	\bm R=\bm x_B-\bm x_A,\qquad R=|\bm R|,
\end{equation}
and the light quark coordinates measured from the two heavy sources are
\begin{equation}
	\bm\xi=\bm x_1-\bm x_A,\qquad
	\bm\eta=\bm x_2-\bm x_B.
\end{equation}
The geometry of the open charm-bottom $c\bar{b}q\bar{q}$ tetraquark system is illustrated in Fig.~\ref{fig:vector_diagram}.

Given that the masses of heavy quarks are much larger than those of light quarks, $m_c, m_b \gg m_q$, the heavy coordinates evolve on a parametrically slower time scale than the light degrees of freedom.
At fixed $\bm R$, the total wave function is therefore written in the adiabatic form
\begin{equation}
	\Psi(\bm R,\bm\xi,\bm\eta)
	=\chi(\bm R)f_{\bm R}(\bm\xi,\bm\eta),
\end{equation}
where $f_{\bm R}(\bm\xi,\bm\eta)$ is the light-field wave function for a fixed heavy-quark separation. 
It satisfies
\begin{equation}
	\left[ \sum_{\text{light}} \frac{p_i^2}{2m_i} 
	+ V_I(\bm R; \bm\xi,\bm\eta) \right] f_{\bm R}(\bm\xi,\bm\eta)
	=\mathcal E(\bm R)f_{\bm R}(\bm\xi,\bm\eta),
	\label{eq4}
\end{equation}
where $V_I(\bm R; \bm\xi,\bm\eta)$ describes the general interactions involving light-light and heavy-light quarks.
At leading adiabatic order, the nonadiabatic derivative couplings generated by $\nabla_{\bm R}f_{\bm R}$ are neglected. 
After removing the heavy center-of-mass motion, the relative heavy wave function then obeys
\begin{equation}
	\left[-\frac{\nabla_{\bm R}^{2}}{2\mu_{c\bar b}}
	+V_{\rm BO}(\bm R)\right]\chi(\bm R)
	=E_{\rm BO}\,\chi(\bm R),
	\label{eq:heavyBO}
\end{equation}
with the unequal-mass reduced mass
\begin{equation}
	\mu_{c\bar b}=\frac{m_c m_b}{m_c+m_b},
	\label{eq:reducedmass}
\end{equation}
and the BO potential
\begin{equation}
	V_{\rm BO}(\bm R) = V(\bm R) + \mathcal E(\bm R) + V_{\text{conf}}.
\end{equation}
Here, $V(\bm R)$ is defined as the interaction between the heavy quarks, and $V_{\text{conf}}$ describes the color-confinement term responsible for the formation of a tetraquark bound state.
Furthermore, the introduction of a reduced mass replaces the standard heavy-mass dynamical term, which is applicable only in the symmetric limit.

\subsection{Color structure and the Born-Oppenheimer potential}

To evaluate the BO interaction, we first need to determine the color structure of the open charm-bottom tetraquarks.
Following our previous study~\cite{Kang:2025xqm}, the color states of the quark clusters $c\bar{b}$ and $q\bar{q}$ are both color octets. 
Therefore, the color coupling by which the two quark clusters form a color singlet is
\begin{equation}
	\left|\chi_c\right\rangle
	= \left|(c\bar b)^8(q\bar q)^8\right\rangle^1,
	\label{eq:colorstate}
\end{equation}
where the superscripts indicate color indices.
For any two given quarks, the one-gluon-exchange (OGE) color-Coulomb term is written as
\begin{equation}
	V_{ij}(r_{ij})=
	\lambda_{ij}\frac{\alpha_s(\mu_{ij})}{r_{ij}},
	\qquad
	\lambda_{ij}
	=\langle\chi_c|\bm T_i\cdot\bm T_j|\chi_c\rangle,
	\label{eq:colorcoulomb}
\end{equation}
where $\bm T_i$ is the generator of the $i$-th particle in the SU(3) color representation.
For the state in Eq.~\eqref{eq:colorstate}, SU(3) recoupling can be obtained via Casimir operators.
Thus, the required color factors we calculate are, respectively: $\lambda_{c\bar b}=\lambda_{q\bar q}=1/6$, $\lambda_{cq}=\lambda_{\bar b\bar q}=-1/3$, and $\lambda_{c\bar q}=\lambda_{q\bar b}=-7/6$.
These coefficients determine both the reference heavy-light interactions and the residual perturbing interactions introduced below.

The running coupling constant of the strong interaction can be evaluated at the reduced mass $\mu_{ij}$ of the corresponding pair; its expression is
\begin{equation}
	\alpha_s(\mu)=
	\frac{\alpha_0}
	{\ln\!\left[(\mu^2+\mu_0^2)/\Lambda_0^2\right]},
	\label{eq8}
\end{equation}
where the parameters are consistent with those in Ref.~\cite{Yu:2026jis}. 
They are $\alpha_0 = 2.118$, $\mu_0 = 36.976$ MeV, and $\Lambda_0 = 0.113 $ fm$^{-1}$, respectively.
All mass scales entering running coupling constant Eq.~\eqref{eq8} are understood to be expressed in the same units. 
Because $\mu_{cq}\neq\mu_{\bar b\bar q}$, the two heavy-light subsystems are inequivalent even when the two light constituents have the same flavor.

To construct a tractable variational calculation to $\mathcal E(R)$, we choose as the reference Hamiltonian two noninteracting heavy-light subsystems, $cq$ and $\bar b\bar q$. 
Their color-Coulomb interactions are
\begin{equation}
	V_{cq}(\xi)
	=-\frac{1}{3}\frac{\alpha_s(\mu_{cq})}{\xi},
	\qquad
	V_{\bar b\bar q}(\eta)
	=-\frac{1}{3}\frac{\alpha_s(\mu_{\bar b\bar q})}{\eta}.
	\label{eq:referencepotentials}
\end{equation}
We use normalized radial trial wave functions,
\begin{equation}
	\zeta_A(\xi)
	=\left(\frac{A^3}{\pi}\right)^{1/2}e^{-A\xi},
	\qquad
	\zeta_B(\eta)
	=\left(\frac{B^3}{\pi}\right)^{1/2}e^{-B\eta},
	\label{eq:trialwf}
\end{equation}
with variational parameters $A$ and $B$. 
In the leading BO treatment, the heavy  $c\bar{b}$ sources are fixed in the light-sector $q\bar{q}$ eigen problem. 
After substituting the interaction potential and the trial wave function into the Hamiltonian, the energy expectation values for the $cq$ and $\bar b\bar q$ clusters are expressed as
\begin{equation}
	\begin{aligned}
		\varepsilon_A(A)
		&=\frac{A^2}{2m_q}
		-\frac{1}{3}\alpha_s(\mu_{cq})A,\\
		\varepsilon_B(B)
		&=\frac{B^2}{2m_{\bar q}}
		-\frac{1}{3}\alpha_s(\mu_{\bar b\bar q})B.
	\end{aligned}
	\label{eq:variationalenergies}
\end{equation}
The parameters $A$ and $B$ are determined independently by minimizing the two expectation values. 
The zeroth-order light-sector wave function reads
\begin{equation}
	f_0(\bm\xi,\bm\eta)
	=\zeta_A(\xi)\,\zeta_B(\eta),
	\label{eq20}
\end{equation}
and the corresponding energy is
\begin{equation}
	\mathcal E_0
	=\varepsilon_A(A)_{\rm min}
	+\varepsilon_B(B)_{\rm min}.
	\label{eq:E0}
\end{equation}

The interactions not included in the reference Hamiltonian are treated to first order part in perturbation theory. 
Using the color factors mentioned above, the Hamiltonian corresponding to the residual interaction is
\begin{align}
	H_{p}={}&
	-\frac{7}{6}\alpha_s(\mu_{q\bar b})
	\frac{1}{|\bm x_1-\bm x_B|}
	-\frac{7}{6}\alpha_s(\mu_{c\bar q})
	\frac{1}{|\bm x_2-\bm x_A|}
	\nonumber\\
	&+\frac{1}{6}\alpha_s(\mu_{q\bar q})
	\frac{1}{|\bm x_1-\bm x_2|}.
	\label{eq:Hp}
\end{align}
The corresponding energy correction is given by
\begin{equation}
	\Delta\mathcal E(\bm R)
	=\langle f_0|H_p|f_0\rangle,
	\label{eq:DeltaEdef}
\end{equation}
which can be written as
\begin{equation}
	\begin{aligned}
		\Delta\mathcal E(\bm R)={}&
		-\frac{7}{6}\alpha_s(\mu_{q\bar b})I_1(R)
		-\frac{7}{6}\alpha_s(\mu_{c\bar q})I_2(R)\\
		&+\frac{1}{6}\alpha_s(\mu_{q\bar q})I_3(R).
	\end{aligned}
	\label{eq:DeltaE}
\end{equation}
Here
\begin{align}
	I_1(R)
	&=\int d^3\xi\,|\zeta_A(\xi)|^2
	\frac{1}{|\bm\xi-\bm R|},\\
	I_2(R)
	&=\int d^3\eta\,|\zeta_B(\eta)|^2
	\frac{1}{|\bm\eta+\bm R|},\\
	I_3(R)
	&=\int d^3\xi\,d^3\eta\,
	|\zeta_A(\xi)|^2|\zeta_B(\eta)|^2
	\frac{1}{|\bm\xi-\bm\eta-\bm R|}.
	\label{eq:Idefinitions}
\end{align}
For the exponential trial wave functions of Eq.~\eqref{eq:trialwf}, the first two integrals are written as
\begin{equation}
	I_1(R)=\frac{1}{R}
	\left[1-(1+AR)e^{-2AR}\right],
	\label{eq:I1}
\end{equation}
\begin{equation}
	I_2(R)=\frac{1}{R}
	\left[1-(1+BR)e^{-2BR}\right].
	\label{eq:I2}
\end{equation}
For $A\neq B$, it is convenient to introduce
\begin{equation}
	\Delta_{AB}\equiv A^2-B^2
\end{equation}
and
\begin{align}
	F_A(R)&=B^4\left(B^2-3A^2-AR\Delta_{AB}\right)e^{-2AR},\\
	F_B(R)&=A^4\left(3B^2-A^2-BR\Delta_{AB}\right)e^{-2BR}.
\end{align}
The light-light color-Coulomb integral is then
\begin{equation}
	I_3(R)=\frac{1}{R}
	\left[1+\frac{F_A(R)+F_B(R)}{\Delta_{AB}^{3}}\right].
	\label{eq:I3}
\end{equation}
It is evident that Eq.~\eqref{eq:I3} exhibits the required asymptotic behavior $I_3(R) \to 1/R$ as $R \to \infty$.
And its apparently singular form at $A=B$ is removable. 
Taking the analytic limit into account, we can obtain
\begin{equation}
	\begin{aligned}
		I_3(R)\big|_{A=B}=\frac{1}{R}\Bigg\{1
		-e^{-2AR}\Bigg[&1+\frac{11}{8}AR
		+\frac{3}{4}(AR)^2\\
		&+\frac{1}{6}(AR)^3\Bigg]\Bigg\}.
	\end{aligned}
	\label{eq:I3equal}
\end{equation}
The adiabatic light-sector energy is therefore
\begin{equation}
	\mathcal E(\bm R)=\mathcal E_0+\Delta\mathcal E(\bm R).
	\label{eq27}
\end{equation}

The direct interaction between the heavy sources is repulsive in the octet $c\bar b$ channel,
\begin{equation}
	V_{c\bar b}(\bm R)
	=\frac{1}{6}\frac{\alpha_s(\mu_{c\bar b})}{R}.
	\label{eq:heavycoulomb}
\end{equation}
Following the phenomenological prescription adopted in our previous study, we supplement the short-distance color interaction by a long-distance confinement term
\begin{equation}
	V_{\rm conf}
	=\sigma \times (R-R_0) \times \theta(R-R_0),
	\label{eq28}
\end{equation}
where $\sigma$ is the linear string tension, and $R_0$ is the phenomenological onset scale of the linear term.
In the present work, $R_0$ is inherited from the calibration adopted in our previous study; its numerical value and the associated sensitivity of the spectrum are discussed in Sec.~\ref{S3}. 
Thus, the resulting BO potential entering Eq.~\eqref{eq:heavyBO} can be expressed as
\begin{equation}
	\begin{aligned}
		V_{\rm BO}(\bm R)
		=\frac{1}{6}\frac{\alpha_s(\mu_{c\bar b})}{R}
		\, &+\mathcal E_0
		+\Delta\mathcal E(\bm R)\\
		&+\sigma \times (R-R_0) \times \theta(R-R_0).
	\end{aligned}
		\label{eq30}	
\end{equation}

Without loss of generality, in the limit of equal heavy-quark masses $m_c = m_b$ and equal orbital parameters $A = B$, the aforementioned equations naturally reduce to the symmetric construction used in our previous study.

\subsection{Wave function basis vectors}

The BO potential introduced above is a color-singlet tetraquark constructed from a color octet-octet configuration.
To evaluate the spin-dependent hyperfine interaction, it is convenient to couple the four constituent spins according to the heavy-light pairing $(cq)(\bar b\bar q)$ employed in the variational ansatz. We emphasize that this choice defines only the spin-coupling scheme; it does not imply that the two heavy-light pairs are separately in pure color representations. 
In fact, the octet-octet color state can be recoupled into the diquark-antidiquark basis as
\begin{equation}
	|(c\bar b)^8(q\bar q)^8\rangle^1
	=-\sqrt{\frac{2}{3}}
	|(cq)^{\bar 3}(\bar b\bar q)^3\rangle^1
	+\frac{1}{\sqrt{3}}
	|(cq)^6(\bar b\bar q)^{\bar 6}\rangle^1.
	\label{eq:diquarkrecoupling}
\end{equation}
Thus, the BO color configuration contains both $\bar{3}\otimes 3$ and $6\otimes\bar{6}$ components and should not be identified with a pure diquark-antidiquark color state.

Since only $S$-wave tetraquarks are considered, the orbital angular momentum is $L=0$ and the parity is positive. 
The total spin can therefore take $J=0$, $1$, or $2$. 
For $J^P=0^+$, we adopt the two spin basis vectors
\begin{equation}
	\begin{aligned}
		|\alpha_1\rangle
		&= |(cq)_{0}(\bar b\bar q)_{0}\rangle_{0},\\
		|\alpha_2\rangle
		&= |(cq)_{1}(\bar b\bar q)_{1}\rangle_{0},
	\end{aligned}
	\label{eq:spinbasis0}
\end{equation}
where the subscripts denote spin indices.
For $J^P=1^+$, the corresponding basis reads
\begin{equation}
	\begin{aligned}
		|\beta_1\rangle
		&= |(cq)_{0}(\bar b\bar q)_{1}\rangle_{1},\\
		|\beta_2\rangle
		&= |(cq)_{1}(\bar b\bar q)_{0}\rangle_{1},\\
		|\beta_3\rangle
		&= |(cq)_{1}(\bar b\bar q)_{1}\rangle_{1},
	\end{aligned}
	\label{eq:spinbasis1}
\end{equation}
whereas for $J^P=2^+$ there is a single basis 
\begin{equation}
	|\gamma_1\rangle
	= |(cq)_{1}(\bar b\bar q)_{1}\rangle_{2}.
	\label{eq:spinbasis2}
\end{equation}

By substituting the above basis vectors into the Hamiltonian \eqref{eq2}, we can obtain the color-spin hyperfine matrices for the corresponding quantum numbers.
These matrices are then diagonalized to determine the spin-dependent mass shifts together with the corresponding normalized eigenvectors.
In principle, a tetraquark state with given quantum numbers can be written as
\begin{equation}
	|\psi \rangle=\sum_\nu c_\nu|g_\nu\rangle,
	\label{eq:spineigenstate}
\end{equation}
where $|g_\nu\rangle$ denotes the appropriate basis defined above and $c_\nu$ is the corresponding eigenvector component.

For the analysis of strong two-body decays, the spin eigenvectors must be recoupled to physical meson-meson channels rather than identified directly with the basis coefficients.
For the rearrangement $(cq)(\bar b\bar q)\to(c\bar q)(\bar b q)$, the spin recoupling coefficient is
\begin{equation}
	\begin{split}
		&\left\langle
		[(s_c s_{\bar q})S_{c\bar q}(s_{\bar b}s_q)S_{\bar b q}]JM
		\middle|
		[(s_c s_q)S_{cq}(s_{\bar b}s_{\bar q})S_{\bar b\bar q}]JM
		\right\rangle
		\\
		&\qquad = \omega\,
		\hat S_{cq}\hat S_{\bar b\bar q}
		\hat S_{c\bar q}\hat S_{\bar b q}
		\begin{Bmatrix}
			s_c & s_q & S_{cq}\\
			s_{\bar q} & s_{\bar b} & S_{\bar b\bar q}\\
			S_{c\bar q} & S_{\bar b q} & J
		\end{Bmatrix},
	\end{split}
	\label{eq:spinrecoupling}
\end{equation}
where the phase factor $\omega=(-1)^{1-S_{\bar b q}-S_{\bar b\bar q}}$ and $\hat S = \sqrt{2S+1}$. 
Besides, a consistent ordering convention for the coupled spinors is understood throughout.

The corresponding color recoupling into the physical meson-meson configuration is
\begin{equation}
	|(c\bar b)^8(q\bar q)^8\rangle^1
	=\frac{2\sqrt{2}}{3}
	|(c\bar q)^1(q\bar b)^1\rangle^1
	-\frac{1}{3}
	|(c\bar q)^8(q\bar b)^8\rangle^1.
	\label{eq:mesonfierz}
\end{equation}
Clearly, the separation amplitude into the $(c\bar q)(q\bar b)$-type channels are determined by the color and spin recoupling coefficients of the configuration, rather than solely by the components $c_i$. 
The color-singlet configuration $(c\bar b)(q\bar q)$, associated with a $c\bar b$ pair plus a light meson, is orthogonal to the pure octet-octet BO structure of Eq.~\eqref{eq:colorstate}. 
Such channels therefore require color-configuration mixing and are not included at leading order within the single-channel approximation adopted here.

\subsection{Partial decay widths}

To predict the strong decays of the open charm-bottom tetraquarks, we consider their dynamically allowed two-body meson-meson channels.
For a parent state of mass $M$ decaying into mesons with masses $m_1$ and $m_2$, the partial width for a two-body $L$-wave decay is \cite{Weng:2019ynv,Guo:2022crh}

\begin{equation}
	\label{eq34}
	\Gamma_\mu = \gamma_\mu \alpha \,\frac{k^{2L+1}}{M^{2L}} \, |c_\mu|^2,
\end{equation}
where the center-of-mass momentum of either daughter meson is expressed as
\begin{align}
	\label{eq:two_body_momentum}
	k = \frac{\sqrt{ \left[M^2 - (m_1 + m_2)^2\right] \left[M^2 - (m_1 - m_2)^2\right] }}{2 M}.
\end{align}

The strong decay channel is closed for $M\leq m_{1}+m_{2}$, in which case its on-shell two-body partial width vanishes. 
The coefficients $\gamma_i$ and $\alpha$ are physical quantities related to the spatial structure, coupling strength, and decay modes. 
Treating this factor as common to different decay channels is a phenomenological approximation rather than a consequence of the BO framework.

The coefficient $c_\mu$ is the overlap with the meson-meson color-spin channel, not an individual component $c_\nu$ of the tetraquark eigenvector in the $(cq)(\bar b\bar q)$ spin basis.
For the spin conserving $S$-wave rearrangement channels, we obtain the transformation relation as follows
\begin{equation}
	\label{eq:recoupled_decay_amplitude}
	c_\mu =\mathcal{C}\,\sum_\nu U_{\mu\nu} c_\nu,
\end{equation}
where $\mathcal{C}$ is an overlap factor for the transformation between different basis vectors.
The spin recoupling coefficient is defined as
\begin{equation}
	\label{eq:spin_overlap_definition}
	U_{\mu\nu}=\big\langle [(s_c s_{\bar q})S_{1\mu}(s_{\bar b}s_q)S_{2\mu}]JM \big|g_\nu\big\rangle.
\end{equation}
Clearly, decays involving higher relative partial waves generally require additional spin-orbit recoupling and transition matrix elements, and are not determined solely by the formula above. 
For two-body decay processes involving tetraquark states with the same configuration, these parameters can be approximately treated as identical.
In this work, we determine the value $\gamma_i \alpha = 0.0126$ by fitting to the classic decay channel $ X(3872) \to J/\psi\rho $.
It should be noted that the spectral function of the $\rho$ is broad. 
When considering only strong $S$-wave decays, we take the mass of the $\rho$ to be 765 MeV.
We can then proceed to estimate the partial decay widths for the predicted open charmed-bottom tetraquarks.

\begin{table}[b]
	\caption{Effective hyperfine couplings (in MeV) extracted from color singlet quark-antiquark ($\kappa_{ij}^{(1)}$) and color antitriplet quark-quark ($\kappa_{ij}^{(\bar{3})}$) systems. 
		In the actual calculations, we take $\kappa_{ij}=\kappa_{ji}$ for a given constituent pair.}
	\label{tab:kappa_parameters}
	\begin{ruledtabular}
		\begin{tabular}{ccccccccc}
			Mesons & $n\bar n$ & $n\bar s$ & $s\bar s$ & $n\bar c$ & $s\bar c$ & $n\bar b$ & $s\bar b$ & $c\bar b$ \\
			$\kappa_{ij}^{(1)}$ & 315 & 195 & 121 & 70 & 72 & 23 & 24 & 20 \\
			\hline
			Baryons & $nn$ & $ns$ & $ss$ & $nc$ & $sc$ & $nb$ & $sb$ & $cb$ \\
			$\kappa_{ij}^{(\bar{3})}$ & 103 & 64 & 72 & 22 & 25 & 6.6 & 7.5 & 10
		\end{tabular}
	\end{ruledtabular}
\end{table}

\section{Results and discussion   \label{S3}}

\subsection{Model parameters}

The model parameters are determined following the phenomenological procedure established in our previous study~\cite{Kang:2025xqm}.
By fitting the conventional hadron spectra—incorporating experimental measurements, available lattice QCD calculations, and other theoretical results—the effective masses of constituent quarks are extracted.
The values are: $m_n=308$ MeV, $m_s=484$ MeV, $m_c=1667$ MeV, and $m_b=5005$ MeV, where $n$ denotes a nonstrange light quark, either $u$ or $d$.
The effective hyperfine couplings are summarized in Table~\ref{tab:kappa_parameters}.
Based on the Casimir scaling hypothesis, for the $|(cq)^{\bar{3}}(\bar{b}\bar{q})^{3}\rangle^{1}$ tetraquark configuration, we have $\kappa_{ij}^{(8)}=\frac{1}{4}\kappa_{ij}^{(1)}$.
It should be emphasized that these parameters are subject to a certain degree of error from the fitting process. 
Therefore, in our calculations, we have accounted for a $\pm20\%$ variation in $\kappa_{ij}$ around its central value.
And this range is a phenomenological variation, not a statistical confidence interval.
The confinement onset parameter $R_0=10$ GeV$^{-1}$ is retained from the previous calibration based on the assumed $X(3872)$ tetraquark assignment. 
These parameters are subsequently used as inputs for the present $c\bar bq\bar q$ calculations without an additional fit to the open heavy flavor tetraquark spectrum.

\begin{figure*}[htbp]
		\centering
	\begin{minipage}{0.33\textwidth}
			\centering
			\includegraphics[width=0.96\textwidth,height=0.96\textwidth]{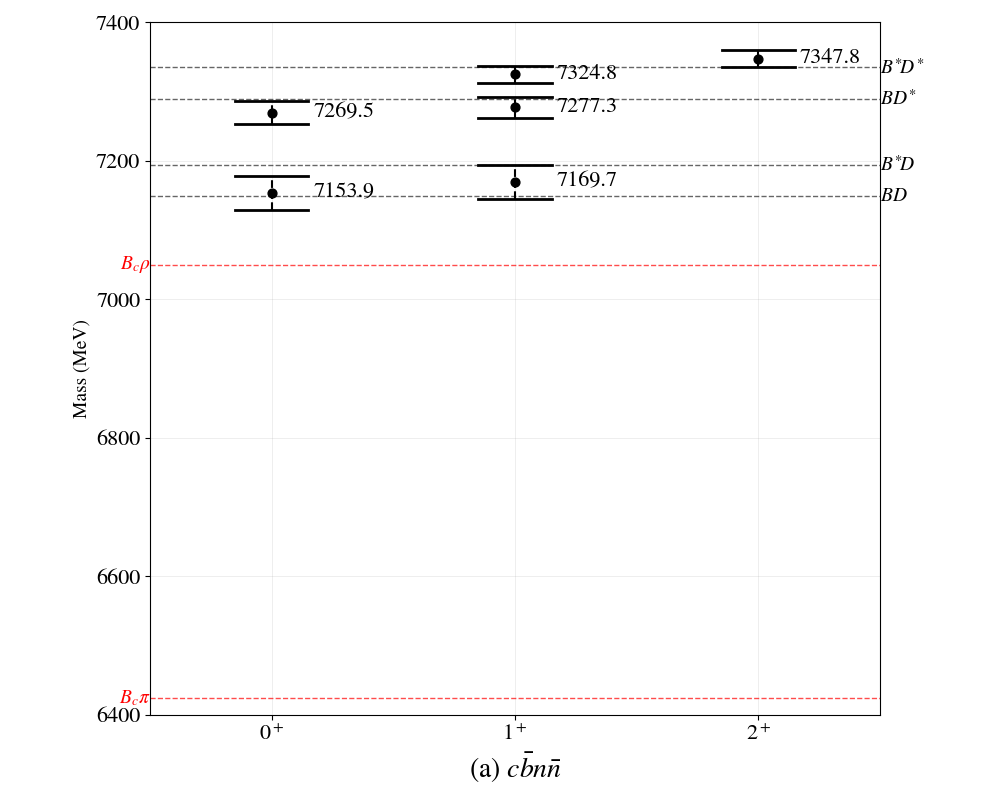}
			
	\end{minipage}\hfill
	\begin{minipage}{0.33\textwidth}
			\centering
			\includegraphics[width=0.96\textwidth,height=0.96\textwidth]{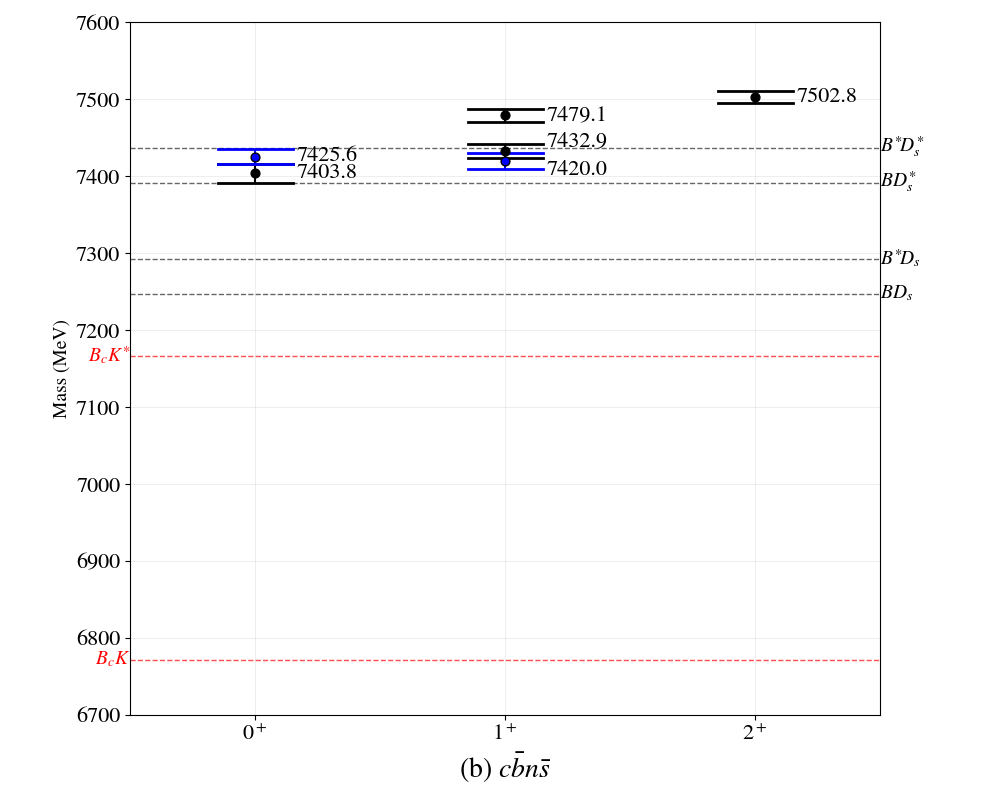}
			
	\end{minipage}\hfill
	\begin{minipage}{0.33\textwidth}
			\centering
			\includegraphics[width=0.96\textwidth,height=0.96\textwidth]{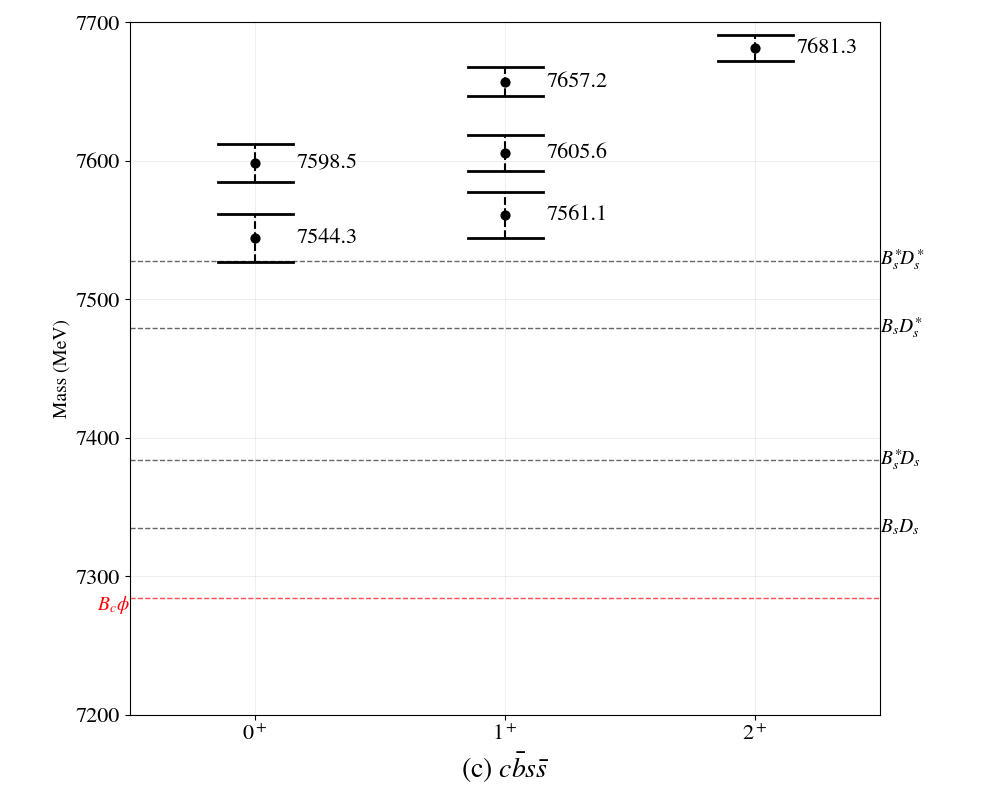}
	\end{minipage}
		\caption{The mass spectra of the $ S $-wave open charm-bottom tetraquark states with distinct quantum numbers: (a) $ c\bar{b}n\bar{n} $, (b) $ c\bar{b}n\bar{s} $, and (c) $ c\bar{b}s\bar{s} $. The black dashed lines and red dashed lines represent the corresponding meson-meson thresholds. The blue spectra are intended to distinguish between states that are very close to one another.}
		\label{fig:three_images_1}
\end{figure*}

\begin{table}[htbp]
	\centering
	\caption{Mass spectra (in MeV) and partial decay widths (in MeV) of the $c\bar{b}n\bar{n}$ tetraquark states.}
	\label{tab:cbnn_decaywidths}
	\footnotesize
	\setlength{\tabcolsep}{3pt}
	\begin{ruledtabular}
		\begin{tabular}{ccccccccc}
			System & $J^P$ & Mass 
			& $BD$
			& $BD^*$
			& $B^*D$
			& $B^*D^*$
			& $B_c\pi$
			& $B_c\rho$ \\
			\hline
			$c\bar{b}n\bar{n}$
			& $0^+$
			& 7269.5 & 1.77 &   &   &   & 2.78 &  \\
			& & 7153.9 & 0.14 &   &   &   & 0.96 &  \\
			& $1^+$
			& 7324.8 &   & 0.01 & 1.72 &   &   & 0.001 \\
			& & 7277.3 &   &   & 0.12 &   &   & 1.60 \\
			& & 7169.7 &   &   &   &   &   & 0.44 \\
			& $2^+$
			& 7347.8 &   &   &   & 0.84 &   &  \\
		\end{tabular}
	\end{ruledtabular}
\end{table}

\begin{table}[htbp]
	\centering
	\caption{Mass spectra (in MeV) and partial decay widths (in MeV) of the $c\bar{b}n\bar{s}$ tetraquark states.}
	\label{tab:cbns_decaywidths}
	\footnotesize
	\setlength{\tabcolsep}{3pt}
	\begin{ruledtabular}
		\begin{tabular}{ccccccccc}
			System & $J^P$ & Mass
			& $BD_s$
			& $BD_s^*$
			& $B^*D_s$
			& $B^*D_s^*$
			& $B_cK$
			& $B_cK^*$ \\
			\hline
			$c\bar{b}n\bar{s}$
			& $0^+$
			& 7425.6 & 1.52 &   &   &   & 1.97 &  \\
			& & 7403.8 & 1.42 &   &   &   & 1.93 &  \\
			& $1^+$
			& 7479.1 &   & 0.02 & 2.00 & 0.55 &   & 0.02 \\
			& & 7432.9 &   & 2.04 & 0.16 &   &   & 2.16 \\
			& & 7420.0 &   & 0.59 & 0.81 &   &   & 0.64 \\
			& $2^+$
			& 7502.8 &   &   &   & 1.90 &   &  \\
		\end{tabular}
	\end{ruledtabular}
\end{table}

\begin{table}[htbp]
	\centering
	\caption{Mass spectra (in MeV) and partial decay widths (in MeV) of the $c\bar{b}s\bar{s}$ tetraquark states.}
	\label{tab:cbss_decaywidths}
	\footnotesize
	\setlength{\tabcolsep}{3pt}
	\begin{ruledtabular}
		\begin{tabular}{ccccccccc}
			System & $J^P$ & Mass
			& $B_sD_s$
			& $B_sD_s^*$
			& $B_s^*D_s$
			& $B_s^*D_s^*$
			& $B_c\phi$ \\
			\hline
			$c\bar{b}s\bar{s}$
			& $0^+$
			& 7598.5 & 2.33 &   &   & 0.76 &  \\
			& & 7544.3 & 1.27 &   &   & 0.60 &  \\
			& $1^+$
			& 7657.2 &   & 0.015 & 2.50 & 0.90 & 0.013 \\
			& & 7605.6 &   & 2.67 & 0.27 & 0.46 & 2.31 \\
			& & 7561.1 &   & 1.01 & 0.82 & 0.59 & 0.88 \\
			& $2^+$
			& 7681.3 &   &   &   & 2.90 &  \\
		\end{tabular}
	\end{ruledtabular}
\end{table}
	
\begin{table*}[htbp]
		\centering
		\caption{Comparison of theoretical predictions for the mass spectra of open bottom-charm tetraquarks (in GeV).}
		\label{tab:tetraquark_mass_comparison}
		\renewcommand{\arraystretch}{1.2}
		\begin{tabular*}{\linewidth}{@{\extracolsep{\fill}}ccccccccc}
			\hline
			\hline  %
			$J^{PC}$ &
			$\vert n_1^{2s_1+1}l_{1j_1}, n_2^{2s_2+1}l_{2j_2}, N^{2s_3+1}L_J \rangle$ & 
			Tetraquark & 
			This work & 
			RQM \cite{Ebert:2007rn} & 
			CMI \cite{Wu:2018xdi} & 
			QCDSR \cite{Zhang:2009vs} & 
			QCDtpSR \cite{Agaev:2016dsg} & 
			RTR \cite{Chen:2025fyh} \\
			\hline  
			$0^{+}$ & $\vert 1^1s_0, 1^1s_0, 1^1S_0 \rangle$ & $(cn)(\bar{b}\bar{n})$ & $7.129\sim7.179$ & 7.177 & 7.152 & 7.12 & 7.11 & 7.17 \\
			&  & $(cn)(\bar{b}\bar{s})$ & $7.392\sim7.416$ & 7.282 & 7.433 &  &  & 7.28 \\
			&  & $(cs)(\bar{b}\bar{s})$ & $7.527\sim7.561$ & 7.398 & 7.502 &  & 7.16 & 7.39 \\
			$0^{+}$ & $\vert 1^3s_1, 1^3s_1, 1^1S_0 \rangle$ & $(cn)(\bar{b}\bar{n})$ & $7.253\sim7.286$ & 7.221 & 7.354 & 7.29 &  &  \\
			&  & $(cn)(\bar{b}\bar{s})$ & $7.416\sim7.435$ & 7.325 & 7.528 &  &  &  \\
			&  & $(cs)(\bar{b}\bar{s})$ & $7.585\sim7.612$ & 7.445 & 7.586 &  &  &  \\
			$1^{+}$ & $\vert 1^1s_0, 1^3s_1, 1^3S_1 \rangle$ & $(cn)(\bar{b}\bar{n})$ & $7.145\sim7.194$ & 7.198 & 7.189 & 7.21 &  & 7.28 \\
			&  & $(cn)(\bar{b}\bar{s})$ & $7.409\sim7.431$ & 7.302 & 7.390 &  &  & 7.39 \\
			&  & $(cs)(\bar{b}\bar{s})$ & $7.545\sim7.578$ & 7.418 & 7.491 &  &  & 7.48 \\
			$1^{+}$ & $\vert 1^3s_1, 1^1s_0, 1^3S_1 \rangle$ & $(cn)(\bar{b}\bar{n})$ & $7.262\sim7.293$ & 7.242 & 7.283 & 7.28 &  & 7.20 \\
			&  & $(cn)(\bar{b}\bar{s})$ & $7.423\sim7.443$ & 7.346 & 7.418 &  &  & 7.32 \\
			&  & $(cs)(\bar{b}\bar{s})$ & $7.593\sim7.619$ & 7.465 & 7.528 &  &  & 7.42 \\
			$1^{+}$ & $\vert 1^3s_1, 1^3s_1, 1^3S_1 \rangle$ & $(cn)(\bar{b}\bar{n})$ & $7.312\sim7.337$ & 7.242 & 7.343 &  &  &  \\
			&  & $(cn)(\bar{b}\bar{s})$ & $7.471\sim7.487$ & 7.345 & 7.460 &  &  &  \\
			&  & $(cs)(\bar{b}\bar{s})$ & $7.647\sim7.667$ & 7.465 & 7.562 &  &  &  \\
			$2^{+}$ & $\vert 1^3s_1, 1^3s_1, 1^5S_2 \rangle$ & $(cn)(\bar{b}\bar{n})$ & $7.336\sim7.360$ & 7.288 & 7.311 &  &  &  \\
			&  & $(cn)(\bar{b}\bar{s})$ & $7.495\sim7.510$ & 7.389 & 7.421 &  &  &  \\
			&  & $(cs)(\bar{b}\bar{s})$ & $7.672\sim7.691$ & 7.506 & 7.554 &  &  &  \\
			\hline
			\hline  
		\end{tabular*}
\end{table*}

\subsection{Spectra and strong decays of the $c\bar{b}q\bar{q}$ tetraquarks}

By properly accounting for the BO approximation and color-spin interactions, we calculate the mass spectra and decays of $S$-wave open charm-bottom $c\bar{b}q\bar{q}$ tetraquarks.
The spectra of the $c\bar{b}n\bar{n}$, $c\bar{b}n\bar{s}$, and $c\bar{b}s\bar{s}$ systems for different quantum numbers are shown in Fig.~\ref{fig:three_images_1}.
Similarly, the corresponding final-state meson-meson thresholds are also plotted in the figure.
The calculated partial decay widths in the kinematically allowed two-body channels are summarized in Tables~\ref{tab:cbnn_decaywidths}--\ref{tab:cbss_decaywidths}. 
In Table~\ref{tab:tetraquark_mass_comparison}, we consider the variation of the color-coupling correlation coefficients $\kappa_{ij}$ and provide the resulting error ranges for the mass spectra, alongside results from other theoretical calculations.
These results help us examine and predict potential open charm-bottom tetraquark states within the corresponding decay channels.

\subsubsection{$c\bar{b}n\bar{n}$ configuration}

As shown in Fig.~\ref{fig:three_images_1}(a), the six predicted $c\bar{b}n\bar{n}$ states occupy the mass region of 7153.9–7347.8 MeV.
For the quantum numbers $0^+$, the two eigenstates are located at 7153.9 and 7269.5 MeV, respectively.
Both are above the $BD$ threshold but below the $B^*D^*$ threshold. 
Taking into account parity conservation in strong interactions, we list in Table \ref{tab:cbnn_decaywidths} the allowed $S$-wave open-flavor decay channels into $BD$ mesons.
The corresponding partial widths are 0.14 and 1.77 MeV, respectively. 
The lower state is especially close to the $BD$ threshold, making the availability and phase space of this decay sensitive to shifts in its predicted mass.
The table also reports $B_c\pi$ estimates of 0.96 and 2.78 MeV; these should be interpreted subject to the color-mixing qualification given above.

Three states with quantum number assignments $1^+$ are obtained at 7169.7, 7277.3, and 7324.8 MeV, respectively.
The lowest lies below the $BD^*$, $B^*D$, and $B^*D^*$ thresholds, and consequently has no on-shell $S$-wave open-flavor decay channels among those considered here. 
Although Table~\ref{tab:cbnn_decaywidths} contains a $B_c\rho$ estimate of 0.44 MeV for this state, that estimate is not a leading-order prediction of the pure octet-octet model.
At 7277.3 MeV, the $B^*D$ channel becomes accessible, with a calculated partial width of 0.12 MeV; $BD^*$ and $B^*D^*$ remain closed. 
The larger tabulated $B_c\rho$ width of 1.60 MeV is again conditional on the additional transition dynamics. 
For the highest $1^+$ tetraquark state, the $B^*D$ channel has the largest calculated rearrangement width, $\Gamma(B^*D)=$1.72 MeV, whereas $\Gamma(BD^*)=$0.01 MeV. 
The strongly unequal strengths of these two open channels provide a more specific decay signature than the mass alone. 
The $B^*D^*$ channel remains kinematically closed at 7324.8 MeV.

The single $2^+$ state is predicted at 7347.8 MeV, above the $B^*D^*$ threshold. 
Its calculated $S$-wave partial width into $B^*D^*$ mesons is 0.84 MeV.
Therefore, among the open-flavor $S$-wave rearrangement decay channels investigated in this work, the $B^*D^*$ final state serves as a characteristic channel for this tensor state.
Higher relative partial waves are not included in the tabulated widths.

\subsubsection{$c\bar{b}n\bar{s}$ configuration}

The replacement of the nonstrange light antiquark by a strange antiquark shifts the predicted levels upward, as seen in Fig.~\ref{fig:three_images_1}(b). 
The six $c\bar{b}n\bar{s}$ states lie between $7403.8$ and $7502.8~\mathrm{MeV}$.
Their masses and the partial widths are collected in Table~\ref{tab:cbns_decaywidths}.

There are two $0^+$ states here, at 7403.8 and 7425.6 MeV, which lie above the $BD_s$ threshold but below the $B^*D_s^*$ threshold.
Their predicted $BD_s$ partial widths are 1.42 and 1.52 MeV, respectively. 
The two scalar levels are separated by only 21.8 MeV, substantially less than the corresponding splitting in the tetraquark $c\bar{b}n\bar{n}$ system. 
The additional $B_c K$ decay channels, with partial widths of 1.93 and 1.97 MeV, respectively. 
However, these are conditional estimates, rather than leading-order open-flavor rearrangement widths.

For the quantum numbers $1^+$, the predicted masses are 7420.0, 7432.9, and 7479.1 MeV.
The two lower states lie below the $B^*D_s^*$ threshold, but decays into $BD_s^*$ and $B^*D_s$ are both kinematically allowed.
Their relative strengths differ markedly: at 7420.0 MeV, the partial widths are 0.59 MeV for $BD_s^*$ and 0.81 MeV for $B^*D_s$, whereas at 7432.9 MeV they are 2.04 and 0.16 MeV, respectively. 
Thus, the two nearby axial-vector states may be distinguished through their contrasting $BD_s^*$ and $B^*D_s$ strengths. 
For the highest state, at 7479.1 MeV, the $B^*D_s$ decay channel has the largest reported rearrangement width, 2.00 MeV, followed by $B^*D_s^*$ at 0.55 MeV; the $BD_s^*$ width is only 0.02 MeV. 
The $B_c K^*$ decay channel estimates in Table~\ref{tab:cbns_decaywidths} require the additional color-transition dynamics discussed above.

The $2^+$ state is found at 7502.8 MeV, above the $B^*D_s^*$ threshold. 
Its calculated $S$-wave partial width in this channel is 1.90 MeV. 
This provides a natural final state in which to examine the predicted tensor level.

\subsubsection{$c\bar{b}s\bar{s}$ configuration}

As displayed in Fig.~\ref{fig:three_images_1}(c), the mass spectrum of the tetraquark $c\bar{b}s\bar{s}$ system occupies a relatively high mass range, from 7544.3 to 7681.3 MeV.
In addition, we present the partial widths for the corresponding decay channels in Table~\ref{tab:cbss_decaywidths}.
Unlike the two lighter flavor configurations, both scalar levels, at 7544.3 and 7598.5 MeV, lie above the $B_s^*D_s^*$ threshold. 
Consequently, both $B_sD_s$ and $B_s^*D_s^*$ are accessible as $S$-wave rearrangement channels. 
For the lower scalar, the respective calculated widths are 1.27 and 0.60 MeV; for the higher scalar they are 2.33 and 0.76 MeV. 
In both cases, $B_sD_s$ is the stronger of the two channels considered.

There are likewise three states with quantum numbers $1^+$, having masses of 7561.1, 7605.6, and 7657.2 MeV, respectively.
For the lowest level, all three tabulated open-flavor rearrangement channels have comparable partial widths: 1.01 MeV for $B_sD_s^*$, 0.82 MeV for $B_s^*D_s$, and 0.59 MeV for $B_s^*D_s^*$. 
No single rearrangement mode overwhelmingly dominates within this set. 
In contrast, the state at 7605.6 MeV favors $B_sD_s^*$, with a partial width of 2.67 MeV, compared with 0.27 and 0.46 MeV for $B_s^*D_s$ and $B_s^*D_s^*$, respectively. 
The highest $1^+$ state has the opposite hierarchy: $B_s^*D_s$ is largest, with $\Gamma(B_s^*D_s)=$2.50 MeV, followed by $B_s^*D_s^*$ at 0.90 MeV, while $\Gamma(B_sD_s^*)=$0.015 MeV. 
These distinct decay patterns offer complementary ways of separating the three axial-vector levels. 
The $B_c\phi$ partial widths also included in Table~\ref{tab:cbss_decaywidths} are subject to the same color-configuration constraints as other $B_c$ decay channels involving a light meson.

Finally, for the quantum number $2^+$, there exists a unique state with a mass of 7681.3 MeV, located above the $B_s^*D_s^*$ threshold. 
The $S$-wave partial width in this channel is calculated to be 2.90 MeV, which is the largest among the partial widths of the three $B^*D^*$-type tensor states considered in the present calculation.

\subsubsection{Comparison and implications}

We are now in a position to provide a comprehensive assessment of the calculation results for the open charm-bottom $c\bar{b}q\bar{q}$ tetraquarks.
Several trends emerge from Fig.~\ref{fig:three_images_1} and Tables~\ref{tab:cbnn_decaywidths}-\ref{tab:cbss_decaywidths}. 
First, the spectra shift to higher masses with increasing strange quark content, although the spin splittings are not uniform across the three flavor sectors. 
For example, the two $0^+$ levels are separated by 115.6, 21.8, and 54.2 MeV in the $c\bar{b}n\bar{n}$, $c\bar{b}n\bar{s}$, and $c\bar{b}s\bar{s}$ systems, respectively. 
Second, the decay modes are controlled jointly by threshold positions and channel-dependent recoupling amplitudes: nearby states with the same $J^P$ can favor different vector-pseudoscalar final states, as illustrated by the $1^+$ levels in the strange sectors. 
Third, the relatively small partial widths quoted here should not be interpreted as the narrow total resonance widths. 
The reason for this is that the current calculations do not account for higher partial waves, additional hadronic channels, coupled-channel effects, and threshold-related dynamics.

Table~\ref{tab:tetraquark_mass_comparison} compares our mass estimates with calculations based on other theoretical approaches.
These approaches primarily include the relativistic quark model (RQM), the chromomagnetic interaction model (CMI), QCD sum rules (QCDSR), the QCD two-point sum rules (QCDtpSR), and Regge trajectory relations (RTR).
Our predictions are generally distributed within the same charm-bottom mass region, although the quantitative differences depend on the flavor contents and spin assignments.
For instance, for the tetraquark $c\bar{b}s\bar{s}$ system with quantum numbers $2^+$, the present central value is $7.681~\mathrm{GeV}$, compared with 7.506 GeV in Ref.~\cite{Ebert:2007rn} and 7.554 GeV in Ref.~\cite{Wu:2018xdi}.
Moreover, assigning levels to individual spin-coupling basis vectors is model dependent when hyperfine mixing is present; comparisons are most meaningful when made between physical eigenstates with the same quantum numbers. 
The error ranges presented in our results reflect the phenomenological variations in the hyperfine coupling, rather than confidence intervals derived from statistical calibration. 
Besides, these errors do not by themselves quantify the systematic uncertainties associated with the BO ansatz or the assumed scale $R_0$ at which confinement begins.

Therefore, the most direct targets for future experiments searching for these open charm-bottom $c\bar{b}q\bar{q}$ tetraquarks are to examine the invariant mass spectra of final states consisting of open flavor pairs, such as $BD$, $B^{*}D^{*}_s$, $B^{*}D^{*}_s$, and $B_s^{*}D_s^{*}$.
Compared to matching a single mass, establishing the decay modes, relative strengths, and priority ordering for several final states would provide more discriminating tests for the unequal-mass BO framework.
A quantitative prediction for the $c\bar{b}$ plus light meson channels will require an extension that includes the relevant color-transition dynamics.

\section{summary \label{S4}}

We have investigated the low-lying $S$-wave $c\bar{b}q\bar{q}$ systems within the BO framework, extending our previous analysis of hidden heavy flavor systems to include unequal-mass heavy quark pairs.
The charm and antibottom sources are treated with the appropriate reduced mass, while the light-sector wave function is constructed from independently optimized heavy-light orbitals.
Starting from a color singlet formed by octet-octet configuration, we combined the BO potential with the color-spin hyperfine interaction to obtain mass spectra with quantum number assignments of $0^+$, $1^+$, and $2^+$. 
The parameters required for the calculation are primarily derived from fitting the conventional hadrons, following a methodology consistent with our earlier work.

For each of the $c\bar{b}n\bar{n}$, $c\bar{b}n\bar{s}$, and $c\bar{b}s\bar{s}$ configurations, we obtain two $0^+$, three $1^+$, and one $2^+$ states. 
Their predicted masses span 7.154-7.348, 7.404-7.503, and 7.544-7.681 GeV, respectively. 
The upward shift with increasing strange quark content is accompanied by flavor-dependent hyperfine splittings.
In particular, the two $0^+$ levels are separated by 115.6, 21.8, and 54.2 MeV in these three sectors. 
Within the adopted phenomenological decay prescription, the calculated partial widths of selected $S$-wave open-flavor rearrangement channels exhibit distinctive spin-dependent patterns. 
The $2^+$ states, for example, have predicted partial widths of 0.84, 1.90, and 2.90 MeV into $B^*D^*$, $B^*D_s^*$, and $B_s^*D_s^*$, respectively. 
Several $1^+$ states also show pronounced differences between the corresponding pseudoscalar-vector and vector-pseudoscalar channels. 
These channel-related partial widths should not be interpreted as the complete total widths or branching fractions, though they do hold some reference value.

Our results are generally consistent with those from other theoretical calculations, with the exception that the masses of tetraquarks containing strange quark components are slightly larger.
Therefore, the mass ordering and relative strengths of the open-flavor final states offer complementary observables for testing the framework. 
Further work incorporating higher partial waves, coupled-channel dynamics, and systematic parameter variations will be necessary to assess the robustness of the mass and decay predictions. 
The results provide quantitative benchmarks for future theoretical studies and experimental searches for open charm-bottom tetraquarks.

\begin{acknowledgments}
	The authors would like to thank Lei Liu for helpful comments and discussions. This work was supported by the Scientific Research Foundation of Chengdu University of Technology under grant No.10912-KYQD2025-09557.
\end{acknowledgments}

\bibliography{ref}

@article{Godfrey:1985xj,
	author = "Godfrey, S. and Isgur, Nathan",
	title = "{Mesons in a Relativized Quark Model with Chromodynamics}",
	doi = "10.1103/PhysRevD.32.189",
	journal = "Phys. Rev. D",
	volume = "32",
	pages = "189--231",
	year = "1985"
}

@article{Capstick:1986ter,
	author = "Capstick, Simon and Isgur, Nathan",
	title = "{Baryons in a relativized quark model with chromodynamics}",
	doi = "10.1103/physrevd.34.2809",
	journal = "Phys. Rev. D",
	volume = "34",
	number = "9",
	pages = "2809--2835",
	year = "1986"
}

@article{Gell-Mann:1964ewy,
	author = "Gell-Mann, Murray",
	title = "{A Schematic Model of Baryons and Mesons}",
	doi = "10.1016/S0031-9163(64)92001-3",
	journal = "Phys. Lett.",
	volume = "8",
	pages = "214--215",
	year = "1964"
}

@article{Belle:2003nnu,
	author = "Choi, S. K. and others",
	collaboration = "Belle",
	title = "{Observation of a narrow charmonium-like state in exclusive $B^\pm \to K^\pm \pi^+ \pi^- J/\psi$ decays}",
	eprint = "hep-ex/0309032",
	archivePrefix = "arXiv",
	doi = "10.1103/PhysRevLett.91.262001",
	journal = "Phys. Rev. Lett.",
	volume = "91",
	pages = "262001",
	year = "2003"
}

@article{Brambilla:2019esw,
	author = "Brambilla, Nora and Eidelman, Simon and Hanhart, Christoph and Nefediev, Alexey and Shen, Cheng-Ping and Thomas, Christopher E. and Vairo, Antonio and Yuan, Chang-Zheng",
	title = "{The $XYZ$ states: experimental and theoretical status and perspectives}",
	eprint = "1907.07583",
	archivePrefix = "arXiv",
	primaryClass = "hep-ex",
	reportNumber = "TUM-EFT 125/19",
	doi = "10.1016/j.physrep.2020.05.001",
	journal = "Phys. Rept.",
	volume = "873",
	pages = "1--154",
	year = "2020"
}

@article{Chen:2022asf,
	author = "Chen, Hua-Xing and Chen, Wei and Liu, Xiang and Liu, Yan-Rui and Zhu, Shi-Lin",
	title = "{An updated review of the new hadron states}",
	eprint = "2204.02649",
	archivePrefix = "arXiv",
	primaryClass = "hep-ph",
	doi = "10.1088/1361-6633/aca3b6",
	journal = "Rept. Prog. Phys.",
	volume = "86",
	number = "2",
	pages = "026201",
	year = "2023"
}

@article{ParticleDataGroup:2024cfk,
	author = "Navas, S. and others",
	collaboration = "Particle Data Group",
	title = "{Review of particle physics}",
	doi = "10.1103/PhysRevD.110.030001",
	journal = "Phys. Rev. D",
	volume = "110",
	number = "3",
	pages = "030001",
	year = "2024"
}

@article{Liu:2024uxn,
	author = "Liu, Ming-Zhu and Pan, Ya-Wen and Liu, Zhi-Wei and Wu, Tian-Wei and Lu, Jun-Xu and Geng, Li-Sheng",
	title = "{Three ways to decipher the nature of exotic hadrons: Multiplets, three-body hadronic molecules, and correlation functions}",
	eprint = "2404.06399",
	archivePrefix = "arXiv",
	primaryClass = "hep-ph",
	doi = "10.1016/j.physrep.2024.12.001",
	journal = "Phys. Rept.",
	volume = "1108",
	pages = "1--108",
	year = "2025"
}

@article{Janssen:1994wn,
	author = "Janssen, G. and Pearce, B. C. and Holinde, K. and Speth, J.",
	title = "{On the structure of the scalar mesons f0 (975) and a0 (980)}",
	eprint = "nucl-th/9411021",
	archivePrefix = "arXiv",
	reportNumber = "KFA-IKP-TH-1994-40, ADP-94-23-T-163",
	doi = "10.1103/PhysRevD.52.2690",
	journal = "Phys. Rev. D",
	volume = "52",
	pages = "2690--2700",
	year = "1995"
}

@article{Shmatikov:1995aj,
	author = "Shmatikov, M.",
	title = "{Hadronic molecules: Meson - baryon hybrids}",
	eprint = "hep-ph/9501259",
	archivePrefix = "arXiv",
	reportNumber = "RUB-MEP-7-95",
	doi = "10.1016/0370-2693(95)00287-U",
	journal = "Phys. Lett. B",
	volume = "349",
	pages = "411--415",
	year = "1995"
}

@article{Guo:2017jvc,
	author = "Guo, Feng-Kun and Hanhart, Christoph and Mei{\ss}ner, Ulf-G. and Wang, Qian and Zhao, Qiang and Zou, Bing-Song",
	title = "{Hadronic molecules}",
	eprint = "1705.00141",
	archivePrefix = "arXiv",
	primaryClass = "hep-ph",
	doi = "10.1103/RevModPhys.90.015004",
	journal = "Rev. Mod. Phys.",
	volume = "90",
	number = "1",
	pages = "015004",
	year = "2018",
	note = "[Erratum: Rev.Mod.Phys. 94, 029901 (2022)]"
}

@article{Jaffe:1976ig,
	author = "Jaffe, Robert L.",
	title = "{Multi-Quark Hadrons. 1. The Phenomenology of $Q^2\bar{Q}^2$ Mesons}",
	reportNumber = "SLAC-PUB-1772",
	doi = "10.1103/PhysRevD.15.267",
	journal = "Phys. Rev. D",
	volume = "15",
	pages = "267",
	year = "1977"
}

@article{Esposito:2016noz,
	author = "Esposito, A. and Pilloni, A. and Polosa, A. D.",
	title = "{Multiquark Resonances}",
	eprint = "1611.07920",
	archivePrefix = "arXiv",
	primaryClass = "hep-ph",
	reportNumber = "JLAB-THY-16-2301",
	doi = "10.1016/j.physrep.2016.11.002",
	journal = "Phys. Rept.",
	volume = "668",
	pages = "1--97",
	year = "2017"
}

@article{Ali:2017jda,
	author = {Ali, Ahmed and Lange, Jens S{\"o}ren and Stone, Sheldon},
	title = "{Exotics: Heavy Pentaquarks and Tetraquarks}",
	eprint = "1706.00610",
	archivePrefix = "arXiv",
	primaryClass = "hep-ph",
	reportNumber = "DESY-17-071",
	doi = "10.1016/j.ppnp.2017.08.003",
	journal = "Prog. Part. Nucl. Phys.",
	volume = "97",
	pages = "123--198",
	year = "2017"
}

@article{Guo:2021yws,
	author = "Guo, Tao and Li, Jianing and Zhao, Jiaxing and He, Lianyi",
	title = "{Mass spectra of doubly heavy tetraquarks in an improved chromomagnetic interaction model}",
	eprint = "2108.10462",
	archivePrefix = "arXiv",
	primaryClass = "hep-ph",
	doi = "10.1103/PhysRevD.105.014021",
	journal = "Phys. Rev. D",
	volume = "105",
	number = "1",
	pages = "014021",
	year = "2022"
}

@article{Xia:2025mgk,
	author = "Xia, Xi and Guo, Tao",
	title = "{Fully heavy tetraquark states with a diquark-antidiquark configuration}",
	eprint = "2508.19838",
	archivePrefix = "arXiv",
	primaryClass = "hep-ph",
	doi = "10.1103/pvq6-kbxf",
	journal = "Phys. Rev. D",
	volume = "113",
	number = "5",
	pages = "054017",
	year = "2026"
}

@article{Close:1998mj,
	author = "Close, F. E. and Page, P. R.",
	title = "{Glueballs}",
	doi = "10.1038/scientificamerican1198-80",
	journal = "Sci. Am.",
	volume = "279",
	pages = "52--57",
	year = "1998"
}

@article{Brunner:2015yha,
	author = {Br{\"u}nner, Frederic and Rebhan, Anton},
	title = "{Nonchiral enhancement of scalar glueball decay in the Witten-Sakai-Sugimoto model}",
	eprint = "1504.05815",
	archivePrefix = "arXiv",
	primaryClass = "hep-ph",
	doi = "10.1103/PhysRevLett.115.131601",
	journal = "Phys. Rev. Lett.",
	volume = "115",
	number = "13",
	pages = "131601",
	year = "2015"
}

@article{Fariborz:2025ybo,
	author = "Fariborz, Amir H.",
	title = "{Spinless glueballs in generalized linear sigma model}",
	eprint = "2508.09474",
	archivePrefix = "arXiv",
	primaryClass = "hep-ph",
	doi = "10.1103/kllv-gj2b",
	journal = "Phys. Rev. D",
	volume = "112",
	number = "11",
	pages = "114007",
	year = "2025"
}

@article{Chanowitz:1982qj,
	author = "Chanowitz, Michael S. and Sharpe, Stephen R.",
	title = "{Hybrids: Mixed States of Quarks and Gluons}",
	reportNumber = "LBL-14865",
	doi = "10.1016/0550-3213(83)90635-1",
	journal = "Nucl. Phys. B",
	volume = "222",
	pages = "211--244",
	year = "1983",
	note = "[Erratum: Nucl.Phys.B 228, 588--588 (1983)]"
}

@article{Klempt:2007cp,
	author = "Klempt, Eberhard and Zaitsev, Alexander",
	title = "{Glueballs, Hybrids, Multiquarks. Experimental facts versus QCD inspired concepts}",
	eprint = "0708.4016",
	archivePrefix = "arXiv",
	primaryClass = "hep-ph",
	doi = "10.1016/j.physrep.2007.07.006",
	journal = "Phys. Rept.",
	volume = "454",
	pages = "1--202",
	year = "2007"
}

@article{Meyer:2015eta,
	author = "Meyer, C. A. and Swanson, E. S.",
	title = "{Hybrid Mesons}",
	eprint = "1502.07276",
	archivePrefix = "arXiv",
	primaryClass = "hep-ph",
	doi = "10.1016/j.ppnp.2015.03.001",
	journal = "Prog. Part. Nucl. Phys.",
	volume = "82",
	pages = "21--58",
	year = "2015"
}

@article{Kang:2025xqm,
	author = "Kang, Bowen and Xia, Xi and Guo, Tao",
	title = "{Hidden heavy flavor tetraquarks in the Born-Oppenheimer approximation}",
	eprint = "2503.10173",
	archivePrefix = "arXiv",
	primaryClass = "hep-ph",
	doi = "10.1103/xnbg-wklj",
	journal = "Phys. Rev. D",
	volume = "111",
	number = "11",
	pages = "114016",
	year = "2025"
}

@article{Piszczatowski2009,
	author = "K. Piszczatowski. and others",
	title = "{Theoretical Determination of the Dissociation Energy of Molecular Hydrogen}",
	doi = "10.1021/ct900391p",
	journal = "J. Chem. Theory Comput.",
	volume = "5",
	number = "3",
	pages = "3039",
	year = "2009"
}

@article{Puchalski:2018pdf,
	author = "Puchalski, Mariusz and Komasa, Jacek and Czachorowski, Pawe{\l} and Pachucki, Krzysztof",
	title = "{Nonadiabatic QED correction to the dissociation energy of the hydrogen molecule}",
	eprint = "1812.02980",
	archivePrefix = "arXiv",
	primaryClass = "physics.atom-ph",
	doi = "10.1103/PhysRevLett.122.103003",
	journal = "Phys. Rev. Lett.",
	volume = "122",
	number = "10",
	pages = "103003",
	year = "2019"
}

@article{Silkowski:2023mjy,
	author = "Si{\l}kowski, Micha{\l} and Komasa, Jacek and Puchalski, Mariusz and Pachucki, Krzysztof",
	title = "{Leading-order QED effects in the ground electronic state of molecular hydrogen}",
	eprint = "2301.07481",
	archivePrefix = "arXiv",
	primaryClass = "physics.atom-ph",
	doi = "10.1103/PhysRevA.107.032807",
	journal = "Phys. Rev. A",
	volume = "107",
	number = "3",
	pages = "032807",
	year = "2023"
}

@article{Fleck:1989mb,
	author = "Fleck, S. and Richard, J. M.",
	title = "{Baryons with double charm}",
	doi = "10.1143/PTP.82.760",
	journal = "Prog. Theor. Phys.",
	volume = "82",
	pages = "760--774",
	year = "1989"
}

@article{Braaten:2014qka,
	author = "Braaten, Eric and Langmack, Christian and Smith, D. Hudson",
	title = "{Born-Oppenheimer Approximation for the XYZ Mesons}",
	eprint = "1402.0438",
	archivePrefix = "arXiv",
	primaryClass = "hep-ph",
	doi = "10.1103/PhysRevD.90.014044",
	journal = "Phys. Rev. D",
	volume = "90",
	number = "1",
	pages = "014044",
	year = "2014"
}

@article{Brambilla:2017uyf,
	author = "Brambilla, Nora and Krein, Gast{\~a}o and Tarr{\'u}s Castell{\`a}, Jaume and Vairo, Antonio",
	title = "{Born-Oppenheimer approximation in an effective field theory language}",
	eprint = "1707.09647",
	archivePrefix = "arXiv",
	primaryClass = "hep-ph",
	reportNumber = "TUM-EFT-69-15",
	doi = "10.1103/PhysRevD.97.016016",
	journal = "Phys. Rev. D",
	volume = "97",
	number = "1",
	pages = "016016",
	year = "2018"
}

@article{Bicudo:2017szl,
	author = "Bicudo, Pedro and Cardoso, Marco and Peters, Antje and Pflaumer, Martin and Wagner, Marc",
	title = "{$u d \bar{b} \bar{b}$ tetraquark resonances with lattice QCD potentials and the Born-Oppenheimer approximation}",
	eprint = "1704.02383",
	archivePrefix = "arXiv",
	primaryClass = "hep-lat",
	doi = "10.1103/PhysRevD.96.054510",
	journal = "Phys. Rev. D",
	volume = "96",
	number = "5",
	pages = "054510",
	year = "2017"
}

@article{Maiani:2019lpu,
	author = "Maiani, Luciano and Polosa, Antonio D. and Riquer, Veronica",
	title = "{Hydrogen bond of QCD in doubly heavy baryons and tetraquarks}",
	eprint = "1908.03244",
	archivePrefix = "arXiv",
	primaryClass = "hep-ph",
	doi = "10.1103/PhysRevD.100.074002",
	journal = "Phys. Rev. D",
	volume = "100",
	number = "7",
	pages = "074002",
	year = "2019"
}

@article{Germani:2020sgm,
	author = "Germani, Davide and Grinstein, Benjamin and Polosa, Antonio Davide",
	title = "{Tetraquarks in the Born-Oppenheimer approximation}",
	eprint = "2501.13249",
	archivePrefix = "arXiv",
	primaryClass = "hep-ph",
	doi = "10.1007/JHEP04(2025)004",
	journal = "JHEP",
	volume = "25",
	pages = "004",
	year = "2020"
}

@article{Berwein:2024ztx,
	author = "Berwein, Matthias and Brambilla, Nora and Mohapatra, Abhishek and Vairo, Antonio",
	title = "{Hybrids, tetraquarks, pentaquarks, doubly heavy baryons, and quarkonia in Born-Oppenheimer effective theory}",
	eprint = "2408.04719",
	archivePrefix = "arXiv",
	primaryClass = "hep-ph",
	reportNumber = "TUM-EFT 185/23",
	doi = "10.1103/PhysRevD.110.094040",
	journal = "Phys. Rev. D",
	volume = "110",
	number = "9",
	pages = "094040",
	year = "2024"
}

@article{Liu:2025jyn,
	author = "Liu, Lei and Xiao, Yanmei and Guo, Tao",
	title = "{Hydrogenlike structures in the strong interaction}",
	eprint = "2505.22177",
	archivePrefix = "arXiv",
	primaryClass = "hep-ph",
	doi = "10.1103/77hs-2gy4",
	journal = "Phys. Rev. D",
	volume = "112",
	number = "5",
	pages = "054021",
	year = "2025"
}

@article{Chen:2016qju,
	author = "Chen, Hua-Xing and Chen, Wei and Liu, Xiang and Zhu, Shi-Lin",
	title = "{The hidden-charm pentaquark and tetraquark states}",
	eprint = "1601.02092",
	archivePrefix = "arXiv",
	primaryClass = "hep-ph",
	doi = "10.1016/j.physrep.2016.05.004",
	journal = "Phys. Rept.",
	volume = "639",
	pages = "1--121",
	year = "2016"
}

@article{Liu:2019zoy,
	author = "Liu, Yan-Rui and Chen, Hua-Xing and Chen, Wei and Liu, Xiang and Zhu, Shi-Lin",
	title = "{Pentaquark and Tetraquark states}",
	eprint = "1903.11976",
	archivePrefix = "arXiv",
	primaryClass = "hep-ph",
	doi = "10.1016/j.ppnp.2019.04.003",
	journal = "Prog. Part. Nucl. Phys.",
	volume = "107",
	pages = "237--320",
	year = "2019"
}

@article{CMS:2025fpt,
	author = "Hayrapetyan, Aram and others",
	collaboration = "CMS",
	title = "{Determination of the spin and parity of all-charm tetraquarks}",
	eprint = "2506.07944",
	archivePrefix = "arXiv",
	primaryClass = "hep-ex",
	reportNumber = "CMS-BPH-24-002, CERN-EP-2025-118",
	doi = "10.1038/s41586-025-09711-7",
	journal = "Nature",
	volume = "648",
	number = "8092",
	pages = "58--63",
	year = "2025"
}

@article{D0:2016mwd,
	author = "Abazov, V. M. and others",
	collaboration = "D0",
	title = "{Evidence for a $B_s^0 \pi^\pm$ state}",
	eprint = "1602.07588",
	archivePrefix = "arXiv",
	primaryClass = "hep-ex",
	reportNumber = "FERMILAB-PUB-16-038-E",
	doi = "10.1103/PhysRevLett.117.022003",
	journal = "Phys. Rev. Lett.",
	volume = "117",
	number = "2",
	pages = "022003",
	year = "2016"
}

@article{ATLAS:2018udc,
	author = "Aaboud, M. and others",
	collaboration = "ATLAS",
	title = "{Search for a Structure in the $B^0_s \pi^\pm$ Invariant Mass Spectrum with the ATLAS Experiment}",
	eprint = "1802.01840",
	archivePrefix = "arXiv",
	primaryClass = "hep-ex",
	reportNumber = "CERN-EP-2017-333",
	doi = "10.1103/PhysRevLett.120.202007",
	journal = "Phys. Rev. Lett.",
	volume = "120",
	number = "20",
	pages = "202007",
	year = "2018"
}

@article{LHCb:2016dxl,
	author = "Aaij, Roel and others",
	collaboration = "LHCb",
	title = "{Search for Structure in the $B_s^0\pi^\pm$ Invariant Mass Spectrum}",
	eprint = "1608.00435",
	archivePrefix = "arXiv",
	primaryClass = "hep-ex",
	reportNumber = "LHCB-PAPER-2016-029, CERN-EP-2016-189",
	doi = "10.1103/PhysRevLett.117.152003",
	journal = "Phys. Rev. Lett.",
	volume = "117",
	number = "15",
	pages = "152003",
	year = "2016",
	note = "[Addendum: Phys.Rev.Lett. 118, 109904 (2017)]"
}

@article{CDF:2017dwr,
	author = "Aaltonen, T. and others",
	collaboration = "CDF",
	title = "{A search for the exotic meson $X(5568)$ with the Collider Detector at Fermilab}",
	eprint = "1712.09620",
	archivePrefix = "arXiv",
	primaryClass = "hep-ex",
	doi = "10.1103/PhysRevLett.120.202006",
	journal = "Phys. Rev. Lett.",
	volume = "120",
	number = "20",
	pages = "202006",
	year = "2018"
}

@article{CMS:2017hfy,
	author = "Sirunyan, A. M. and others",
	collaboration = "CMS",
	title = "{Search for the X(5568) state decaying into $\mathrm{B}^{0}_{\mathrm{s}}\pi^{\pm}$ in proton-proton collisions at $\sqrt{s} = $ 8 TeV}",
	eprint = "1712.06144",
	archivePrefix = "arXiv",
	primaryClass = "hep-ex",
	reportNumber = "CMS-BPH-16-002, CERN-EP-2017-287",
	doi = "10.1103/PhysRevLett.120.202005",
	journal = "Phys. Rev. Lett.",
	volume = "120",
	number = "20",
	pages = "202005",
	year = "2018"
}

@article{LHCb:2020bls,
	author = "Aaij, Roel and others",
	collaboration = "LHCb",
	title = "{A model-independent study of resonant structure in $B^+\to D^+D^-K^+$ decays}",
	eprint = "2009.00025",
	archivePrefix = "arXiv",
	primaryClass = "hep-ex",
	reportNumber = "LHCb-PAPER-2020-024, CERN-EP-2020-158",
	doi = "10.1103/PhysRevLett.125.242001",
	journal = "Phys. Rev. Lett.",
	volume = "125",
	pages = "242001",
	year = "2020"
}

@article{LHCb:2020pxc,
	author = "Aaij, Roel and others",
	collaboration = "LHCb",
	title = "{Amplitude analysis of the $B^+\to D^+D^-K^+$ decay}",
	eprint = "2009.00026",
	archivePrefix = "arXiv",
	primaryClass = "hep-ex",
	reportNumber = "LHCb-PAPER-2020-025, CERN-EP-2020-159",
	doi = "10.1103/PhysRevD.102.112003",
	journal = "Phys. Rev. D",
	volume = "102",
	pages = "112003",
	year = "2020"
}

@article{LHCb:2022sfr,
	author = "Aaij, R. and others",
	collaboration = "LHCb",
	title = "{First Observation of a Doubly Charged Tetraquark and Its Neutral Partner}",
	eprint = "2212.02716",
	archivePrefix = "arXiv",
	primaryClass = "hep-ex",
	reportNumber = "LHCb-PAPER-2022-026, CERN-EP-2022-239",
	doi = "10.1103/PhysRevLett.131.041902",
	journal = "Phys. Rev. Lett.",
	volume = "131",
	number = "4",
	pages = "041902",
	year = "2023"
}

@article{LHCb:2022lzp,
	author = "Aaij, R. and others",
	collaboration = "LHCb",
	title = "{Amplitude analysis of $B^0 \rightarrow \bar{D}^0 D_s^+ \pi^-$ and $B^+ \rightarrow D^- D_s^+ \pi^+$ decays}",
	eprint = "2212.02717",
	archivePrefix = "arXiv",
	primaryClass = "hep-ex",
	reportNumber = "LHCb-PAPER-2022-027, CERN-EP-2022-246",
	doi = "10.1103/PhysRevD.108.012017",
	journal = "Phys. Rev. D",
	volume = "108",
	number = "1",
	pages = "012017",
	year = "2023"
}

@article{Liu:2020nil,
	author = "Liu, Ming-Zhu and Xie, Ju-Jun and Geng, Li-Sheng",
	title = "{$X_0(2866)$ as a $D^*\bar{K}^*$ molecular state}",
	eprint = "2008.07389",
	archivePrefix = "arXiv",
	primaryClass = "hep-ph",
	doi = "10.1103/PhysRevD.102.091502",
	journal = "Phys. Rev. D",
	volume = "102",
	number = "9",
	pages = "091502",
	year = "2020"
}

@article{Zhang:2020oze,
	author = "Zhang, Jian-Rong",
	title = "{Open-charm tetraquark candidate: Note on $X_0$(2900)}",
	eprint = "2008.07295",
	archivePrefix = "arXiv",
	primaryClass = "hep-ph",
	doi = "10.1103/PhysRevD.103.054019",
	journal = "Phys. Rev. D",
	volume = "103",
	number = "5",
	pages = "054019",
	year = "2021"
}

@article{Wang:2020prk,
	author = "Wang, Guang-Juan and Meng, Lu and Xiao, Li-Ye and Oka, Makoto and Zhu, Shi-Lin",
	title = "{Mass spectrum and strong decays of tetraquark ${\bar{c}}{\bar{s}} qq$ states}",
	eprint = "2010.09395",
	archivePrefix = "arXiv",
	primaryClass = "hep-ph",
	doi = "10.1140/epjc/s10052-021-08978-0",
	journal = "Eur. Phys. J. C",
	volume = "81",
	number = "2",
	pages = "188",
	year = "2021"
}

@article{Guo:2021mja,
	author = "Guo, Tao and Li, Jianing and Zhao, Jiaxing and He, Lianyi",
	title = "{Mass spectra and decays of open-heavy tetraquark states}",
	eprint = "2108.06222",
	archivePrefix = "arXiv",
	primaryClass = "hep-ph",
	doi = "10.1103/PhysRevD.105.054018",
	journal = "Phys. Rev. D",
	volume = "105",
	number = "5",
	pages = "054018",
	year = "2022"
}

@article{Duan:2023qsg,
	author = "Duan, Man-Yu and Wang, En and Chen, Dian-Yong",
	title = "{Searching for the open flavor tetraquark $T_{c\bar{s}0}(2900)^{++}$ in the process $B^+\rightarrow K^+ D^+ D^-$}",
	eprint = "2305.09436",
	archivePrefix = "arXiv",
	primaryClass = "hep-ph",
	doi = "10.1140/epjc/s10052-024-13044-6",
	journal = "Eur. Phys. J. C",
	volume = "84",
	number = "7",
	pages = "681",
	year = "2024"
}

@article{Wang:2023hpp,
	author = "Wang, Bo and Chen, Kan and Meng, Lu and Zhu, Shi-Lin",
	title = "{Spectrum of the molecular tetraquarks: Unraveling the $T_{cs0}(2900)$ and $T^a_{c\bar{s}0}(2900)$}",
	eprint = "2309.02191",
	archivePrefix = "arXiv",
	primaryClass = "hep-ph",
	doi = "10.1103/PhysRevD.109.034027",
	journal = "Phys. Rev. D",
	volume = "109",
	number = "3",
	pages = "034027",
	year = "2024"
}

@article{Ebert:2007rn,
	author = "Ebert, D. and Faustov, R. N. and Galkin, V. O. and Lucha, W.",
	title = "{Masses of tetraquarks with two heavy quarks in the relativistic quark model}",
	eprint = "0706.3853",
	archivePrefix = "arXiv",
	primaryClass = "hep-ph",
	reportNumber = "HU-EP-07-21, HEPHY-PUB-842-07",
	doi = "10.1103/PhysRevD.76.114015",
	journal = "Phys. Rev. D",
	volume = "76",
	pages = "114015",
	year = "2007"
}

@article{Ortega:2020uvc,
	author = "Ortega, Pablo G. and Segovia, Jorge and Entem, David R. and Fernandez, Francisco",
	title = "{Spectroscopy of ${B_c}$ mesons and the possibility of finding exotic ${B_c}$-like structures}",
	eprint = "2001.08093",
	archivePrefix = "arXiv",
	primaryClass = "hep-ph",
	doi = "10.1140/epjc/s10052-020-7764-6",
	journal = "Eur. Phys. J. C",
	volume = "80",
	number = "3",
	pages = "223",
	year = "2020"
}

@article{Wu:2018xdi,
	author = "Wu, Jing and Liu, Xiang and Liu, Yan-Rui and Zhu, Shi-Lin",
	title = "{Systematic studies of charmonium-, bottomonium-, and $B_c$-like tetraquark states}",
	eprint = "1810.06886",
	archivePrefix = "arXiv",
	primaryClass = "hep-ph",
	doi = "10.1103/PhysRevD.99.014037",
	journal = "Phys. Rev. D",
	volume = "99",
	number = "1",
	pages = "014037",
	year = "2019"
}

@article{Chen:2025fyh,
	author = "Chen, Jiao-Kai and Song, He and Liu, Xin-Ru",
	title = "{$\lambda $ and $\rho $ Regge trajectories for bottom-charm tetraquarks $(bq)(\bar{c}\bar{q}')$ and $(cq)(\bar{b}\bar{q}')$}",
	eprint = "2508.18899",
	archivePrefix = "arXiv",
	primaryClass = "hep-ph",
	doi = "10.1140/epjc/s10052-025-15075-z",
	journal = "Eur. Phys. J. C",
	volume = "85",
	number = "11",
	pages = "1344",
	year = "2025"
}

@article{Wang:2019xzt,
	author = "Wang, Zhi-Gang",
	title = "{Analysis of the axialvector $B_c$-like tetraquark states with the QCD sum rules}",
	eprint = "1907.10921",
	archivePrefix = "arXiv",
	primaryClass = "hep-ph",
	doi = "10.1209/0295-5075/128/11001",
	journal = "EPL",
	volume = "128",
	number = "1",
	pages = "11001",
	year = "2019"
}

@article{Agaev:2016dsg,
	author = "Agaev, S. S. and Azizi, K. and Sundu, H.",
	title = "{Open charm-bottom scalar tetraquarks and their strong decays}",
	eprint = "1611.00293",
	archivePrefix = "arXiv",
	primaryClass = "hep-ph",
	doi = "10.1103/PhysRevD.95.034008",
	journal = "Phys. Rev. D",
	volume = "95",
	number = "3",
	pages = "034008",
	year = "2017"
}

@article{Liu:2025fhl,
	author = "Liu, Zhe and Xu, Hao and Liu, Xiang",
	title = "{$D^{(\ast)}\bar{B}^{(\ast)}$ dynamics in chiral effective field theory}",
	eprint = "2503.10299",
	archivePrefix = "arXiv",
	primaryClass = "hep-ph",
	doi = "10.1103/l9pk-hqx1",
	journal = "Phys. Rev. D",
	volume = "112",
	number = "3",
	pages = "034007",
	year = "2025"
}

@online{Yu:2026jis,
	author = "Yu, Yongtao and Guo, Tao",
	title = "{Spectra and nonleptonic decays of multiply heavy baryons in a finite-size quark-diquark model}",
	eprint = "2610.05721",
	archivePrefix = "arXiv",
	primaryClass = "hep-ph",
	month = "10",
	year = "2026"
}

@article{Weng:2019ynv,
	author = "Weng, Xin-Zhen and Chen, Xiao-Lin and Deng, Wei-Zhen and Zhu, Shi-Lin",
	title = "{Hidden-charm pentaquarks and $P_c$ states}",
	eprint = "1904.09891",
	archivePrefix = "arXiv",
	primaryClass = "hep-ph",
	doi = "10.1103/PhysRevD.100.016014",
	journal = "Phys. Rev. D",
	volume = "100",
	number = "1",
	pages = "016014",
	year = "2019"
}

@article{Guo:2022crh,
	author = "Guo, Tao and Li, Jianing and Zhao, Jiaxing and He, Lianyi",
	title = "{Investigation of the tetraquark states $Qq\bar{Q} \bar{q}$ in the improved chromomagnetic interaction model}",
	eprint = "2211.10834",
	archivePrefix = "arXiv",
	primaryClass = "hep-ph",
	doi = "10.1088/1674-1137/accb87",
	journal = "Chin. Phys. C",
	volume = "47",
	number = "6",
	pages = "063107",
	year = "2023"
}

@article{Zhang:2009vs,
	author = "Zhang, Jian-Rong and Huang, Ming-Qiu",
	title = "{$\{Q\bar{q}\} \{\bar{Q}^\prime q\}$ molecular states}",
	eprint = "0906.0090",
	archivePrefix = "arXiv",
	primaryClass = "hep-ph",
	doi = "10.1103/PhysRevD.80.056004",
	journal = "Phys. Rev. D",
	volume = "80",
	pages = "056004",
	year = "2009"
}

\end{document}